\documentclass[prd,twocolumn,aps,amsmath,amssymb,amsfonts,superscriptaddress,nofootinbib]{revtex4}
\usepackage[english]{babel}
\usepackage{bm} 

\newcommand{\beq}{\begin{equation}}
\newcommand{\eeq}{\end{equation}}
\newcommand{\bea}{\begin{eqnarray}}
\newcommand{\eea}{\end{eqnarray}}

\newcommand{\de}{\mbox{${\delta}$}}

\newcommand{\om}{\mbox{${\omega}$}}

\newcommand{\pa}{\mbox{${\partial}$}}

\newcommand{\lsim}{\mbox{$<$\hspace{-0.8em}\raisebox{-0.4em}{$\sim$}}}

\begin{document}
\selectlanguage{english}

\title{Comment on ``Rovibrational quantum interferometers and gravitational waves"}

\author{\firstname{I.~B.}~\surname{Khriplovich}}
\email{khriplovich@inp.nsk.su}
 \affiliation{Budker Institute of
Nuclear Physics, 630090, Novosibirsk, Russia}
\author{\firstname{S.~K.}~\surname{Lamoreaux}}
\email{steve.lamoreaux@yale.edu}
 \affiliation{Yale University, Department of Physics, P.O. Box 208120, New Haven, CT 06520-8120}
\author{\firstname{A.~O.}~\surname{Sushkov}}
\email{alex.sushkov@yale.edu}
 \affiliation{Yale University, Department of Physics, P.O. Box 208120, New Haven, CT 06520-8120}
\author{\firstname{O.~P.}~\surname{Sushkov}}
\email{sushkov@phys.unsw.edu.au} \affiliation{School of Physics,
University of New South Wales, Sydney, NSW 2052, Australia}

\begin{abstract}
In a recent paper, Wicht, L\"{a}mmerzahl, Lorek, and Dittus [Phys. Rev.
{\bf A 78}, 013610 (2008)] come to the conclusion that a molecular
rotational-vibrational quantum interferometer may possess the sensitivity
necessary to detect gravitational waves. We do not agree with their
results and demonstrate here that the true sensitivity of such
interferometer is many orders of magnitude worse than that
claimed in the mentioned paper.
In the present comment we estimate the expected energy shifts
and derive equations of motion for a quantum symmetric top
(diatomic molecule or deformed nucleus) in the field
of gravitational wave, and then estimate the sensitivity
of possible experiments.

\end{abstract}

\maketitle

The direct detection of gravitational waves with an experimental apparatus located on or near the
Earth is one of the forefront goals of modern science.  Several large-scale detector projects,
which are presently being either built or designed, promise the first glimpses of this form
of radiation that most certainly must permeate the fabric of the Universe.
Given the scale of these experiments, it is reasonable to ask whether gravitational
waves might be more simply detected.  In a recent paper, Wicht et al. ~\cite{wicht}
suggest that it is possible to detect such waves by measuring the energy shifts of rovibrational levels of a diatomic molecule.  We show
here that any such effect is many orders of magnitude smaller than
predicted in \cite{wicht}.

Following Ref.~\cite{wicht} we start with the problem of the hydrogen atom in
the field of a weak gravitational wave propagating along the $z$ axis,
$h_{\mu\nu}(t-z)$. The so-called TT gauge, where
\beq\label{g}
h_{11} = - h_{22}, \quad h_{12} = h_{21}, \quad h_{0\mu} =
h_{3\mu}=0\,,
\eeq
is convenient for the description of this wave.
Wicht et al~\cite{wicht} base their analysis on the covariant Klein-Gordon
equation. Expansion of this equation in powers of $h_{ij}$ and $1/c$
($c$ is speed of light)
gives the following nonrelativistic
hamiltonian for the electron of mass $m$
and charge $-e$
\beq\label{ham}
 H = - \frac{\hbar^2}{2m} (\de_{ij} - h_{ij})\pa_i \pa_j - eA_0\,,
\eeq
and where the field of a Coulomb center with charge $e$ in the
presence of the gravitational wave is
\beq\label{A0}
A_0 = \frac{e}{r}\left(1 - \frac{h_{ij}x_i x_j}{2r^2}\right)\,.
\eeq
This is the same hamiltonian used in \cite{wicht}, which we agree is valid.

Let us now calculate the energy shifts generated by the gravitational wave.
The total system hamiltonian is
\begin{eqnarray}
\label{pt}
H&=&H_0+\delta H \nonumber\\
H_0&=&\frac{p^2}{2m}-\frac{e^2}{r} \nonumber\\
\delta H &=& \frac{\hbar^2}{2m}h_{ij}\pa_i \pa_j+\frac{e^2h_{ij}x_i x_j}{2r^3}\ .
\end{eqnarray}
Naively, the energy shift is of the order of the perturbation
\beq
\label{nes}
\delta \epsilon ~ \sim \langle \frac{\hbar^2}{2m}h_{ij}\pa_i \pa_j\rangle
~\sim \langle\frac{e^2h_{ij}x_i x_j}{2r^3}\rangle \sim h \ Ry \ ,
\eeq
where $Ry$ stands for Rydberg, a typical atomic energy.
However, we claim that the shift is much smaller, on the order
of
\beq
\label{es}
\delta \epsilon ~ \sim h\frac{(\hbar \omega)^2}{Ry} \ ,
\eeq
where $\omega$ is the frequency of the gravitational wave, which is typically much lower
than atomic or molecular frequencies of interest.

To prove this we perform the following transformation of coordinates
$x \to y$ in the hamiltonian (\ref{ham}),(\ref{A0}).
\beq
\label{tr}
y_i = x_i + \frac{1}{2} h_{ij} x_j \ .
\eeq
This is the transformation into the frame freely falling in
the gravitational wave.
After this transformation,
the interaction with the gravitational wave is completely removed
from Eqs.  (\ref{ham}) and (\ref{A0}).
However, this does not mean that the interaction with the gravitational
wave vanishes completely.
Consider the time-dependent Schr\"odinger equation
\beq
\label{se1}
i\hbar \left.\frac{\partial}{\partial t}\right|_x\psi(t,x)=H\psi(t,x) \ .
\eeq
The partial time derivative is taken at constant $x$.
Under the transformation in Eq.(\ref{tr}) the left hand side of Eq.(\ref{se1})
is changed as
\beq
\label{ttp}
\left.\frac{\partial}{\partial t}\right|_x =
\left.\frac{\partial}{\partial t}\right|_y
+\frac{\partial y_i}{\partial t}\frac{\partial}{\partial y_i}
=\left.\frac{\partial}{\partial t}\right|_y
+\frac{1}{2}{\dot h}_{ik}\frac{\partial}{\partial y_i}
 \ .
\eeq
Therefore Schr\"odinger Eq. (\ref{se1}) is transformed to
\beq
\label{se2}
i\hbar \frac{\partial}{\partial t}\psi(t,y)=\left[H_0(y)
+\frac{1}{2}{\dot h}_{ik}y_kp_i\right]\psi(t,y) \ ,
\eeq
where the equation is written in the $y$-coordinates and
the time derivative is necessarily taken at $y=\rm{constant}$.
Hence, the new hamiltonian is
\beq
\label{ham2}
H'=
H_0+\frac{1}{2}{\dot h}_{ik}y_kp_i
=
H_0+\frac{1}{2}m{\dot h}_{ik}y_k{\dot y}_i\ ,
\eeq
where we have used $p_i=m{\dot y}_i$.
The new perturbation is
\begin{eqnarray}
\label{ham3}
\delta H'&=&
\frac{1}{2}m{\dot h}_{ik}y_k{\dot y}_i
=\frac{d}{dt}\left(
\frac{1}{4}m{\dot h}_{ik}y_ky_i\right)
-\frac{1}{4}m{\ddot h}_{ik}y_ky_i\nonumber\\
&\to&-\frac{1}{4}m{\ddot h}_{ik}y_ky_i \ .
\end{eqnarray}
In this equation we have thrown away the full time derivative
because it influences neither the equations of motion nor the physical results.
Equation (\ref{ham3}) justifies our estimate (\ref{es}) for the energy
shift or for a typical matrix element, if we substitute ${\ddot h}_{ik}\sim \omega^2 h$ and
$y_i,y_k\sim a_B=\hbar^2/me^2$, where $a_B$ is the Bohr radius.

Obviously, the above statements are valid not only for a hydrogen atom.
One can perform the transformation (\ref{tr}) for any quantum system
in  a gravitational wave; furthermore
one can certainly work in terms of original $x$-coordinates as well.
For example, for the problem of the gravitational radiation of atomic hydrogen,
the equivalence of the two approaches has been demonstrated
explicitly in Ref.~\cite{br}.
However, in the case of low frequency gravitational waves,
calculations in the $x$-coordinates (``$x$-frame'' calculations) are ``dangerous'' because
there is a dramatic exact cancelation (or compensation) between different large terms in the perturbation $\delta H$ given by Eq.
(\ref{pt})
and approximations can easily destroy this exact compensation.
Because the result in \cite{wicht} does not contain the frequency of the gravitational wave
and parametrically coincides with the naive estimate (\ref{nes}),
we believe that this compensation was destroyed in Ref.~\cite{wicht}
due to approximations in the molecular calculations for the ionic molecule HD$^+$.

One can derive Eq. (\ref{ham3}) in a much simpler way using
a frame freely falling in the gravitational wave
from the very beginning.
In this frame the center of mass of the
system remains at rest. The motion of any particle of this system
with respect to the center of mass is described by the well-known
covariant equation (see, e.g., \cite{web,lppt})
\beq
{D^2\eta^\mu \over Ds^2}\,+R^\mu_{\;\;\rho\nu\tau} u^\rho u^\tau
\eta^\nu = 0\,.
\eeq
In the general case, this equation for the so-called geodesic
deviation $\eta^\mu$ has the following meaning: Particles $a$ and
$b$ move along two close geodesics $x^\mu_a(s)$ and $x^\mu_b(s)$;
then, by definition, $\eta^\mu(s) = x^\mu_a(s)-x^\mu_b(s)$, and
$u^\mu = dx^\mu_b(s)/ds$.

In our problem, let the index $b$ label the coordinate of center
of mass which is at rest in the chosen reference frame. Then
$u^\mu = (1,0,0,0)$, and in the weak field of the gravitational
wave $h_{mn} = h_{mn}^{(0)}\, \exp[-i\om (t-z)]$ we arrive at the
following equation of motion for particle $a$:
\beq\label{dd}
\ddot{x}_m^a = - \frac{1}{2}\; \om^2 h_{mn} x_n^a\,.
\eeq

In the hydrogen atom (or hydrogen-like ion), where the coordinates
of electron, center of mass, and proton (or nucleus), $x^e$,
$x^c$, and $x^p$, respectively, lie on a straight line, we arrive
immediately at the following equation of motion for the relative
coordinate $x = x^e - x^p$:
\beq\label{ddh}
\ddot{x}_m = - \frac{1}{2}\; \om^2 h_{mn} x_n\,.
\eeq
The corresponding interaction hamiltonian for the hydrogen atom is
therefore
\beq\label{ddhH}
H_{int} = \frac{1}{4}\; \mu^{ep} \,\om^2 h_{mn} x_m x_n\,;
\eeq
\[
\mu^{ep} = \frac{m^e m^p}{m^e + m^p} \simeq m^e\,.
\]
Let us note here that because $h_{mn}$ depend on $t-z$, but not on
$x_{m,n}$, the relations (\ref{ddh}) and (\ref{ddhH}) are in exact
correspondence.
Eq. (\ref{ddhH}) coincides with Eq. (\ref{ham3}).

The molecular ion HD$^+$ considered in Ref. \cite{wicht} consists
of three constituents: an electron, a proton, and a deuteron. To
simplify the description, we note that, with $m^e \ll m^p$, $m^d$,
one can assume that the coordinates of proton, center of mass, and
deuteron, $x^p$, $x^c$, and $x^d$, respectively, lie on a straight
line, so that for the relative coordinate $x = x^p - x^d$ we have
the same equation, (\ref{ddh}). Therefore, the interaction
hamiltonian for the HD$^+$ ion can be written as
\beq\label{hd}
H_{int} =  \frac{1}{4}\; \mu^{dp}\, \om^2\,h_{mn} x_m x_n\,;
\eeq
\[
\mu^{dp} = \frac{m^d m^p}{m^d + m^p} \simeq \frac{2}{3}\,m^p.
\]
We have neglected here the formally-arising term
\[
\frac{1}{4}\; m^e \om^2 \,h_{mn} x_m^e x_n^e\,,
\]
because the assumption $m^e \ll m^p$ has been made already for the
derivation of hamiltonian (\ref{hd}); furthermore, the contribution of
this term to the discussed effect is relatively suppressed as $m^e/m^p$.

The typical frequency shift for a HD$^+$ ion, as estimated with
hamiltonian in Eq. (\ref{hd}), is
\beq\label{shift}
\de \nu \sim \frac{1}{\hbar}m^p\om^2 h a_B^2.
\eeq
This estimate is valid as long as the frequency $\omega$ of the gravitational
wave is much less than that of the molecular vibrations and rotations.
 For the dimensionless amplitude $h \sim 10^{-19}$ and frequency
$\om \sim 10^4$ Hz, we estimate a frequency shift as $\sim 10^{-23}$ Hz.
This estimate is well below the corresponding
estimate $\sim 3\times 10^{-5}$ Hz presented in Ref.~\cite{wicht}.
Moreover, this frequency shift varies with time together with $h(t-z)$.
With this time-dependence of the frequency shift, the observation of the effect
seems even less realistic because high precision spin precession measurements
generally require considerable averaging time (days to years).


Let us make some additional remarks about the behavior of a molecule (or a deformed nucleus)
under the influence of the hamiltonian (\ref{hd}). This hamiltonian
influences both the vibrational and
rotational degrees of freedom of the diatomic molecule.
The classical (nonquantum) vibrational dynamics in the gravitational wave
are discussed in several textbooks, see e.g. Ref.~\cite{mtw},
and quantum mechanical treatment does not bring significant differences in the dynamics.
Here we derive the implications of Eq. (\ref{hd}) for rotational dynamics.
The classical equation of motion for angular momentum
${\bm L}={\bm r}\times{\bm p}$ follows from Eq. (\ref{hd})
\begin{equation}
\label{jdot}
{\dot L}_i=\frac{1}{2}\omega^2\epsilon_{ikl}h_{ln}I_{kn} \ ,
\end{equation}
where is $I_{kn}$ is the tensor of inertia of the molecule.
Only the traceless part of $I_{kn}$ contributes to Eq. (\ref{jdot}), and
the molecule is a symmetric top.  As usual we denote by $I_{||}$ the
moment of inertia for rotations around axis of the top, and by
$I_{\perp}$ the moment of rotations around a perpendicular
axis.\footnote{For a molecule or deformed nucleus one must
set $I_{||}=0$. Nevertheless, we keep $I_{||}$ to stress
that only traceless part of the inertia tensor contributes to
the spin precession.}
Then
\begin{equation}
\label{ikn}
I_{kn}=\frac{1}{3}\left(2I_{\perp}+I_{||}\right)\delta_{kn}+
\left(I_{||}-I_{\perp}\right)\left(n_kn_n-\frac{1}{3}\delta_{kn}\right).
\end{equation}
Clearly  Eqs.(\ref{jdot}) and (\ref{ikn}) are valid for both
the classical and quantum cases.
Let us consider a molecule in a rotational quantum state with
a given angular momentum $J$, ${\bm L}=\hbar{\bm J}$,
${\hat {\bm J}}^2=J(J+1)$.
Then (see e.g. \cite{LL})
\begin{eqnarray}
\label{iknj}
I_{kn}&\to&{\cal I}\left(J_kJ_n+J_nJ_k-\frac{2}{3}J(J+1)\delta_{kn}\right)
\nonumber\\
{\cal I}&=&\left(I_{||}-I_{\perp}\right)\frac{3\Omega^2-J(J+1)}
{J(2J-1)(J+1)(2J+3)} \ .
\end{eqnarray}
Here we keep only the traceless part of the inertia tensor,
and $\Omega$ is projection of ${\bm J}$ on axis of the top.
Hence Eq. (\ref{jdot}) is transformed to
\begin{equation}
\label{jdot1}
{\dot J}_i=\frac{\cal I}{2\hbar}\omega^2\epsilon_{ikl}h_{ln}
\left\{J_kJ_n+J_nJ_k\right\} \ .
\end{equation}
The corresponding effective Hamiltonian is
\begin{equation}
\label{heff}
H_{eff}= - \frac{\cal I}{2}\omega^2h_{ln}J_lJ_n \ .
\end{equation}
Eqs. (\ref{iknj}) and (\ref{heff}) are equally applicable to
a diatomic molecule or to a deformed nucleus, and describe transitions or energy shifts between magnetic sublevels.
For the ground state $J=\Omega=S$, $S$ is spin of the system.
We assume that $S> 1/2$.
Hence Eq. (\ref{jdot1}) is transformed to equation of spin precession
in the field of gravitational wave
\begin{equation}
\label{sdot1}
{\dot S}_i= - \frac{I_{\perp}-I_{||}}{2\hbar(S+1)(2S+3)}
\omega^2\epsilon_{ikl}h_{ln}(S_kS_n+S_nS_k) \ .
\end{equation}
This is a general equation for a free spin.

Given that a magnetic resonance frequency shift due to (\ref{sdot1}) is
extremely small, we  alternatively consider
the case of a solid, where there is a possibility of a much larger
nuclear electric quadrupole field shift. This shift is induced by the deformation of the
lattice due to the gravitational wave. For a sample with size $\sim 1$ m and for
$\omega \gtrsim 10^4Hz$ the frequency of the gravitational wave
disappears from the equations because the frequency of the lowest acoustic
standing wave $\omega_0$ is lower than $\omega$.
The cancellation of the $\omega$-dependence is similar to that
for the LIGO experiment discussed below, see eqs. (\ref{LIGO2}), (\ref{LIGO3}).
Hence the maximum possible nuclear quadrupole resonance frequency shift is
\begin{equation}
\label{NQR}
{\dot S}\sim 10^9h \ Hz \ .
\end{equation}
for the case of deformed
nuclei.

Unfortunately the frequency shift $10^{-10}$ Hz given by
Eq. (\ref{NQR}) with $h\sim 10^{-19}$ is still very small.
The current state of the art nuclear spin precession frequency measurements
accuracy (for a gas of $^{199}$Hg atoms used in a permanent electric dipole
moment (EDM) search \cite{uwhgedm}) is barely at this level after several
years of measurement and averaging, and this experimental accuracy far
exceeds that of any other experiment performed to date.  Of course, the
possibility of a macroscopic sample with a large number of participating
spins suggests that some increase in sensitivity might be possible, but there
is a considerable loss in sensitivity due to large-sample measurement
limitations. The only plausible measurement technique is the much less
efficient monitoring of the sample's precessing magnetization with a SQUID
or other magnetometer, compared to the direct optical detection of the spin
state that can be used with gaseous samples.

Let us add a brief remark about gravity wave detectors based on strain measurements (such as LIGO).
These are also described by Eq.~(\ref{dd}). Suppose a mass (such as an interferometer mirror)
is positioned in equilibrium at $x=y=z=0$, and another mass (the second mirror)
is positioned in equilibrium at $x=L$, $y=z=0$. The mass at the origin is unaffected by
the gravity wave, the response of the mass at $x=L$ is given by
\begin{equation}
\label{LIGO1}
\ddot{x}+\omega_0^2(x-L)= - \frac{1}{2}\omega^2 h_+(t) L,
\end{equation}
where $\omega_0$ is the natural oscillation frequency around the equilibrium point,
$h_+=h_{11}=-h_{22}$, and we neglect damping. For a wave $h_+(t)=h_+^{(0)}e^{-i\omega t}$
the solution is
\begin{equation}
\label{LIGO2}
x=L+\frac{1}{2}\frac{\omega^2}{\omega^2-\omega_0^2}h_+^{(0)} L e^{-i\omega t}.
\end{equation}
LIGO is a ``free-mass'' detector, which means that $\omega_0\ll\omega$
($\omega_0\approx 4$~Hz for Advanced LIGO \cite{AdvLIGO}). This gives the strain
\begin{equation}
\label{LIGO3}
\Delta L/L=\frac{1}{2} h_+^{(0)} e^{-i\omega t}.
\end{equation}
The strain in oscillator modes with frequencies much greater than the gravitational wave frequency
is suppressed by $\omega^2/\omega_0^2$. Indeed, the more rigid the system, the less sensitive it is to an
external force. In a molecule, for example, $\omega_0$ is on the order of
THz, which gives an enormous strain suppression for  gravitational
wave frequencies $\om \;\lsim \;10^4$ Hz.

\section*{Acknowledgements}
The authors would like to acknowledge useful discussions with David DeMille, Dmitry Budker, and
Max Zolotorev.
The work by I.B. Khriplovich
was supported by the Russian Foundation for Basic Research through
Grant No. 08-02-00960-a.

\end{document}